\documentclass[showpacs,preprintnumbers,amsmath,amssymb,epsfig]{revtex4}
\usepackage{graphicx}
\usepackage{dcolumn}
\usepackage{bm}
\newcommand{\menge}[1]{{\mathchoice{\mbox{\sf #1}}{\mbox{\sf #1}}{\mbox{%
\scriptsize\sf #1}}{\mbox{\scriptsize\sf #1}}}}
\newcommand{\nz}{\ifmmode \menge{I\hskip -.2ex N} \else
                          {\sf I\hskip -.2ex N} \fi}
\begin{document}
\title{Generation of entangled states of two
three-level atoms in cavity QED}
\author{XuBo Zou, K. Pahlke and W. Mathis}
\address{Electromagnetic Theory Group at THT,\\
 Department of Electrical
Engineering, University of Hannover, Germany}

\begin{abstract}
We present a scheme to generate maximally entangled states of two
three-level atoms with a nonresonant cavity by cavity-assisted
collisions. Since the cavity field is only virtually excited no
quantum information will be transferred from the atoms to the
cavity.
\end{abstract}  \maketitle

Entangled states of quantum particles give the possibility to test
quantum mechanics against a local hidden variable theory
\cite{EPR, bell,ghz}. They also have practical applications in
quantum cryptography \cite{cry}, quantum dense coding \cite{qua}
and quantum teleportation \cite{tele}. Most of the research in
quantum information processing is based on entanglement generation
of quantum two-level systems (Qubits), which represent the
information. Recently, there is much interest in $N$-level quantum
systems ($N\ge3$) to represent information. It was demonstrated
that key distributions based on three-level quantum systems are
more secure against eavesdropping than those based on two-level
systems \cite{mo}. Key distribution protocols based on entangled
three-level systems were also proposed \cite{cerf}. The security
of these protocols is related to the violation of the Bell
inequality. Recently, it was shown that the quantum prediction
differs more radically from classical physics in the case of
three-level systems than in the case of two-level systems. The
three-level system provides in this context a much smaller level
of noise \cite{kp}. The proof of Bell's theorem without the
inequalities by Greenberger, Horne, and Zeilinger (GHZ) was
extended to multiparticle multi-level systems \cite{cab}. One way
to generate multiqubit entanglement of $N$-level quantum systems
is to use higher order parametric down conversion \cite{three}. A
more direct way is to use multilevel quantum systems. In this
context entanglement generation of $N$-level quantum systems was
reported \cite{three1,three2}. Rydberg atoms which cross
superconductive cavities are an almost ideal system to generate
entangled states and to perform small scale quantum information
processing \cite{qed}. A number of schemes were proposed in the
context of cavity QED. In particular EPR pairs \cite{epr} and GHZ
states \cite{ghz1} were successfully generated by a successive
interaction of a series of atoms in a cavity field. In these
schemes cavities act as memories, which store the information of
an electric system and transfer it back to the electric system.
Thus, the decoherence of the cavity field becomes one of the main
obstacles for the implementation of quantum information in cavity
QED. Recently, significant progress was made by proposals for
atoms, which interact with a nonresonant cavity \cite{zhang}. In
this theoretical scheme it is suggested to use a dynamic, which
involves a virtual exchange of a photon with the field. To the
first order of the approximation the scheme is insensitive to
cavity losses or to the presence of a stray of a thermal field in
the mode. Recently, this process was also suggested to generate
GHZ states \cite{state} and to implement the quantum search
algorithm \cite{algo}. Following the proposal of Ref\cite{zhang}
an experiment was performed in which two Rydberg atoms cross a
nonresonant cavity. These atoms became entangled in a controllable way
by cavity-assisted collisions \cite{bbb}.\\

In this paper we show that cavity-assisted collisions of two
Rydberg atoms can be used to entangle their electronic states
$|f\rangle$, $|e\rangle$ and $|g\rangle$ in the related
three-level modell (see Fig.1(a)). The experimental setup is shown
in Fig.1(b). The Fabry-Perot resonator, which is denoted by cavity
$C$, sustains a resonant cavity mode of frequency $\omega_a$. The
$|e\rangle\leftrightarrow|g\rangle$ and
$|f\rangle\leftrightarrow|g\rangle$ transitions are at $51.1$ and
$54.3$ GHz, respectively. The cavity mode is shifted in the
frequency from the transitions $|e\rangle\leftrightarrow|g\rangle$
and $|f\rangle\leftrightarrow|g\rangle$ by detunings $\delta_{eg}$
and $\delta_{gf}=\delta_{eg}+\delta_{det}$. The value
$\delta_{det}=3.2$ GHz is the frequency difference of the
transitions $|e\rangle\leftrightarrow|g\rangle$ and
$|f\rangle\leftrightarrow|g\rangle$. The experimental values,
which are given in Ref\cite{qed} show $\delta_{eg}\ll\delta_{gf}$.
Thus, we can choose the cavity frequency in a way that only the
levels $|e\rangle$ and $|g\rangle$ are appropriately affected by
the nonresonant atom-field coupling. The quantum state $|f\rangle$
will in a good approximation not be
affected during the atom-cavity interaction.\\

We write the interaction Hamiltonian inside of the cavity in the
interaction picture
\begin{equation}
H=g[e^{-i\delta_{eg}{t}}a^{\dagger}(\sigma_{1-}+\sigma_{2-})
+e^{i\delta_{eg}{t}}a(\sigma_{1+}+\sigma_{2+})]\,.
\label{1}
\end{equation}
Where $\sigma_{j-}=|g_j\rangle\langle{e_j}|$ and
$\sigma_{j+}=|e_j\rangle\langle{g_j}|$, $a$ and $a^{\dagger}$ are
the annihilation and creation operator of the cavity field. The
atom-cavity coupling strength is denoted by $g$. In the large
detuning case of $\delta_{eg}\gg {g}$ no energy exchange between
the atoms and the cavity will happen. The effective Hamiltonian is
given by \cite{zhang}
\begin{equation}
H=\lambda\sum_{i,j=1}^2(\sigma_{i+}\sigma_{j-}aa^{\dagger}-\sigma_{i-}\sigma_{j+}a^{\dagger}a)\,,
\label{2}
\end{equation}
where $\lambda=g^2/\delta_{eg}$. If the cavity field is at the
beginning in the vacuum state the effective Hamiltonian (\ref{2})
reduces to
\begin{equation}
H=\lambda(\sigma_{1+}\sigma_{1-}+\sigma_{2+}\sigma_{2-}+\sigma_{1+}\sigma_{2-}+\sigma_{2+}\sigma_{1-})
\,.\label{3}
\end{equation}
The first two terms describe the Stark shift in the vacuum cavity.
The dipole coupling between the two atoms, which is induced by the
cavity, are considered by the other terms. In order to generate
maximally entangled states of two three-level atoms, we assume
that two atoms are initially prepared in the state
$|e_1\rangle|e_2\rangle$. The atom $1$ crosses two classical
fields, which are tuned to the transitions
$|e\rangle\leftrightarrow|g\rangle$ and
$|g\rangle\leftrightarrow|f\rangle$, respectively. By choosing the
amplitudes and phases of the classical fields appropriately this
atom becomes prepared in the state
$\sqrt{\frac{1}{3}}|f_1\rangle-\sqrt{\frac{2}{3}}|g_1\rangle$.
Then both atoms are simultaneously sent into the cavity $C$, which
is in the vacuum state. The interaction is described by the
effective Hamiltonian (\ref{3}), which causes no effect on the
state $|f_1\rangle|e_2\rangle$. After the interaction time $t_1$
the quantum state
\begin{equation}
\Psi(t_1)=\sqrt{\frac{1}{3}}|f_1\rangle|e_2\rangle-\sqrt{\frac{2}{3}}e^{-i\lambda
t_1}[\cos(\lambda t_1)|g_1\rangle|e_2\rangle-i\sin(\lambda
t_1)|e_1\rangle|g_2\rangle ] \label{4}
\end{equation} is
obtained. With the choice of $\lambda t_1=\pi/2$ the state
(\ref{4}) becomes
\begin{equation}
\Psi(\pi/2\lambda)=\sqrt{\frac{1}{3}}|f_1\rangle|e_2\rangle
+\sqrt{\frac{2}{3}}|e_1\rangle|g_2\rangle\,. \label{5}
\end{equation}
The atom $2$ is then addressed by a classical microwave pulse,
which is tuned to the transition
$|e\rangle\leftrightarrow|f\rangle$. This step requires separate
addressing of the atoms. Since microwave field can not be focused
narrow enough to address atom $2$ without affecting atom $1$ a
method should be used, which makes the atomic transitions of the
two atoms slightly different. An appropriate method was proposed
by Yamaguchi et al. \cite{algo}, which is based on the Stark
effect. The authors suggest to use a set of electrodes in the
cavity, which create an inhomogeneous electric field. Since the
two atoms are located in space regions with different electric
field strength their frequencies become independently
controllable. The amplitudes and phases of the classical fields
have to be chosen appropriately in order to let the atom $2$
undergo the transition $|e_2\rangle\longrightarrow
|f_2\rangle,|f_2\rangle\longrightarrow-|e_2\rangle $. Here we
assume that the classical microwave field is strong enough so that
the nonlinear interaction between atoms can be neglected during
this stage. After this operation the state (\ref{5}) becomes
\begin{equation}
\Psi^{\prime}(\pi/2\lambda)=\sqrt{\frac{1}{3}}|f_1\rangle
|f_2\rangle+\sqrt{\frac{2}{3}}|e_1\rangle|g_2\rangle\,. \label{6}
\end{equation}
Then the classical microwave field is
switched off and the evolution of the system is determined by the
interaction (\ref{3}). After another interaction time $t_2$ the
system's time evolution has transformed the state (\ref{6}) to the
state
\begin{equation}
\Psi(t_1+t_2)=\sqrt{\frac{1}{3}}|f_1\rangle|f_2\rangle+\sqrt{\frac{2}{3}}e^{-i\lambda
t_2}[\cos(\lambda t_2)|e_1\rangle|g_2\rangle-i\sin(\lambda
t_2)|g_1\rangle|e_2\rangle] \,.\label{7}
\end{equation}
If we choose $\lambda t_2=\pi/4$ the quantum state
\begin{equation}
\Psi(3\pi/4\lambda)=\sqrt{\frac{1}{3}}(|f_1\rangle|f_2\rangle+e^{-i\pi/4}
|e_1\rangle|g_2\rangle-ie^{-i\pi/4}|g_1\rangle|e_2\rangle)
\label{8}
\end{equation}
will be obtained. After the two atoms left the cavity $C$ the atom
$2$ crosses a classical field, which is tuned to the transition
$|e\rangle\leftrightarrow|g\rangle$. If the amplitude and the
phase of the classical field is chosen appropriately the atom $2$
will undergo the transition $
|e_2\rangle\longrightarrow-e^{-i\pi/4}|g_2\rangle,
|g_2\rangle\longrightarrow e^{i\pi/4}|e_2\rangle $. Thus, the
state (\ref{8}) becomes
\begin{equation}
\Psi_{bell}=\sqrt{\frac{1}{3}}(|f_1\rangle|f_2\rangle+
|e_1\rangle|e_2\rangle+|g_1\rangle|g_2\rangle)\,. \label{9}
\end{equation}
This is the maximally entangled state of a two three-level system,
which is discussed in Ref\cite{kp}.\\
One of the difficulties of this scheme is the requirement to sent
two atoms simultaneously through the cavity, otherwise an error
will result. In the following, we discuss the case, that the
second atom enters the cavity in the excited state before the
first atom. This deviation from the ideal case shall be considered
with the time difference $\Delta\tau$, which denotes a fraction of
the Rabi frequency $\tau=\pi/\lambda=\pi\delta_{eg}/g^2$ of the
Hamiltonian (\ref{2}). Then the quantum state
\begin{equation}
\Psi=\sqrt{\frac{1}{3}}|f_1\rangle|f_2\rangle+\sqrt{\frac{2}{3}}e^{i\Delta\pi}
[\cos(\frac{\pi}{4}-\Delta\pi)|e_1\rangle|e_2\rangle+\sin(\frac{\pi}{4}-\Delta\pi)|g_1\rangle|g_2\rangle]
\label{10}
\end{equation}
will be generated. The difference of the state (\ref{10}) to the
state (\ref{9}) can be characterized in terms of the fidelity
$F=|\langle\Psi_{bell}|\Psi\rangle|^2$:
\begin{equation}
F=\frac{5+4\cos2\Delta\pi}{9}\,. \label{11}
\end{equation}
If $\Delta=0.01$ holds, we have $F=0.999$. In this case the
operation is only slightly affected.\\

It is necessary to give a brief discussion on the experimental
realization of the proposed scheme. It was reported that the
cavity can have a photon storage time of $T=1$ms (corresponding to
$Q=3\times10^8$). The radiative time of the Rydberg atoms with the
principle quantum numbers $49$, $50$ and $51$ is about
$3\times10^{-2}$s \cite{qed}. The coupling constant of the atoms
to the cavity field is $g/2\pi=25$kHz \cite{bbb}. In order to
achieve a good entanglement in the cavity-assisted collision
process, the detuning $\delta_{eg}$ should be much bigger than
$g$. With the choice $\delta_{eg}=10g$ the interaction time
between the atom and the cavity field is in the order
$3\pi\delta_{eg}/4g^2\simeq 1.5 \times 10^{-4}$s. The time needed
for the classical field pulse is at this scale negligible. Thus,
the interaction time needed to complement the total procedure is
much shorter than the radiative time and the photon lifetime $1ms$
in the present cavity. For the interaction time
$1.5\times10^{-4}$s the velocity of the prepared atoms should be
$v_p\simeq0.7\times10^{4}L$, where $L$ is the length of the
cavity. If we choose $L=2.75cm$ the velocity of the atoms should
be of the order of $192m/s$, which is in the range of present
experiments. Based on cavity QED techniques the present scheme
seems to become realizable in a near future.\\

In summary, we have proposed a scheme to generate maximally
entangled states of two three-level atoms. During the passage of
the atoms through the cavity field they are only virtually excited.
No transfer of quantum
information will happen between the atoms and the cavity. The
experimental implementation of the scheme demonstrates the power of
cavity QED to manipulate complex entangled states for quantum
information processing.

\begin{flushleft}

{\Large \bf Figure Captions}

\vspace{\baselineskip}

{\bf Figure 1.(a)}  This figure shows the electronic levels of the
three-level atom modell in the energy representation.

{\bf Figure 1.(b)} This figure shows the experimental apparatus.
The atoms $1$ and $2$ cross the cavity with the same velocity but
at different positions with a different electric field strength.
This makes an individual manipulation of each atom by a classical
field possible. Inside the cavity atom $2$ is manipulated by the
classical field $S$. Outside the cavity both atoms are manipulated
by the classical fields $R_1$ and $R_2$, respectively.

\end{flushleft}

\end{document}